\documentclass[prb,groupedaddress,twocolumn,showkeys,showpacs,floatfix]{revtex4}
\usepackage{epsfig}
\usepackage{color}
\usepackage{amssymb,amsmath}
\begin{document}
\title{Partial spin polarization of conductance in vertical bi-layer nanowire with
rectangular and smooth
lateral confinement potential}
\author{T. Chwiej}
\email{chwiej@fis.agh.edu.pl}
\affiliation{AGH University of Science and Technology, al. A. Mickiewicza 30, 30-059 Cracow,
Poland}
\begin{abstract}
\noindent
We simulate the electron transport in vertical bi-layer nanowire which can be fabricated
in molecular beam epitaxy process with lateral confinement potential formed by means of
cleaved overgrowth or surface oxidization methods giving rectangular and smooth side confinement,
respectively. In calculations we take into account interaction between charge carriers using DFT
within local spin density approximation. If magnetic field
is perpendicular to the wire axis, pseudogaps are opened in energy dispersion relation E(k) what in
conjunction with spin Zeeman shift of spin-up and spin-down subbands allows for quite high spin
polarization of conductance.  We find that in nanowire with rectangular lateral confinement
potential, electron density has two maximums localized at wire edges in each layers. This
modificates strongly all magnetosubbands giving up to four energy minimums in lowest subband and
considerably diminishes widths of pseudogaps  what translates into low maximal spin polarization of
conductance, not exceeding $40\%$.
This drawback is absent in wire with smooth lateral confinement. However, in order to gain a
large spin polarization simultaneous tuning of magnetic field as well as the Fermi energies in both
layers of nanowire are required.
\end{abstract}
\maketitle

\section{Introduction}

Influence of magnetic field on electron transport in double-nanowire system  depends on mutual
orientation of magnetic field and wire axis.\cite{thomas,fischer4,lyo3} In longitudinal magnetic
field i.e. directed
parallel to wire axis, the wave functions are compressed in direction being perpendicular to
electron motion what changes the tunnel coupling bewteen layers.\cite{thomas,moroukh}
On the other hand, perpendicular magnetic field may mix vertical or lateral modes leading to
subbands hybrydization.\cite{lyo1,shi} Due to magnetic mixing of wave
functions localized in different layers, crossings of subbands in energy dispersion relation E(k)
are replaced by avoided corissings.  In such case, the pseudogaps are open in energy spectrum what
brings severe consequences for conductance depending on quality of
nanowire.\cite{lyo1,lyo2,chwiej-bilayer}
If transport layers are vertically stacked in nanowire, one over another, and, tilted magnetic field
is applied to a system, then more pseudogaps may appear in energy spectrum since both, inter-layer
as well as intra-layer modes mixing are allowed what makes such mixing to be more
effective.\cite{chwiej-bilayer}
Recently we have indictated\cite{chwiej-bilayer} that spin polarization of conductance in
bi-layer nanowire may reach up to $60\%$ for moderate Fermi energy when only a few transport
channels are open. Those preliminary calculations however have not taken into account an
electrostatic interaction between electrons in nanosystem which may have a great impact on electron
transport\cite{zigzag,drag,manybody,wigner,wigner2} and only rectangular shape of lateral
confinement was considered.

In this work, we continue our theoretical studies of single electron transport in bi-layer nanowire
but now we will mainly focus on spin-polarization of conductance when strong magnetic field
is applied perpendicularily to wire axis and electrostatic interactions bewteen electrons are taken
into account within local spin density approximation. In our considerations we model two types of
side confinement potentials, that is, the lateral barriers will have the rectangular or smooth
shapes. 
First type of confinement can be formed by etching of quantum wire from two-dimensional
nanostructure that holds an electron gas confined within two layers and next cleaving its side
surfaces. This process\cite{cleavededge} removes defects from surfaces and then, after
deposition of barrier material on both sides of nanowire gives finally side barriers without
surface charge. Cross section of confinement potential of such nanowire resembles rectangular
well.\cite{cleavededge}
Second type of the confinement can be formed by surface oxidization of the nanostructure. Two
parallel nanogrooves created during this process may sink up to the depth of 85 nanometers
\cite{groove_depth} with the width of 50 nanometers.\cite{groove_width}  Charge gathered on walls of
each nanogrooves forms
electrostatic barriers  which smoothly deplet electron gas confined in both, the upper and lower
wells in wire. This technique is fast and very flexible since it allows for formation of much more
complicated patterns of confinement potentials like e.g. laterally coupled two quantum rings
\cite{ring_ring} or a ring capacitively coupled with quantum point contact.\cite{ring_qpc}
The shape of lateral confinement in nanowire remarkably influence on
spatial distribution of electron density in quantum wire and 
the overall electrostatics in nanosystem. Thus, it has a strong impact on single electron
transport properties.\cite{ihnatsenka1,ihnatsenka2}

We found that two edge wells are formed in wire when lateral confinement has rectangular shape.
Maxima of electron density are localized in edge wells in each layers what considerably modificates
magnetosubbands. For that reason, widths of pseudogaps are substantially reduced and spin
polarization of conductance does not  exceed $40\%$, provided that, the width of nanowire equals few
tens of nanometers. 
However, if lateral confinement is smooth, the electron densities in both layers have single
elongated maximums and neighbouring magnetosubbands are well separated on energy scale.
The results of our simulations show that an accidental overlapping of spin-up and spin-down energy
branches can decrease maximal value of conductance polarization but we think it can be  
increased if both, the magnetic field and Fermi energy will simultaneously be tuned.

Paper is organized as follows. In Sec.\ref{Sec:theo} we will present details of theoretical
model that was used in calculations,
an influence of rectangular and smooth lateral confinement potentials on
magnetosubbands and spin polarization of conductance of bi-layer nanowire is presented and
disscused in Sec.\ref{Sec:rect} and in Sec.\ref{Sec:smooth}, respectively.
Section \ref{Sec:con} includes conclusions.

\section{Theoretical model}
\label{Sec:theo}

\subsection{Model of nanostructure}
\label{Sec:model}

\begin{figure}[htbp!]
\hbox{
	\epsfxsize=70mm
       \rotatebox{0}{{ \epsfbox[2 7 540 744] {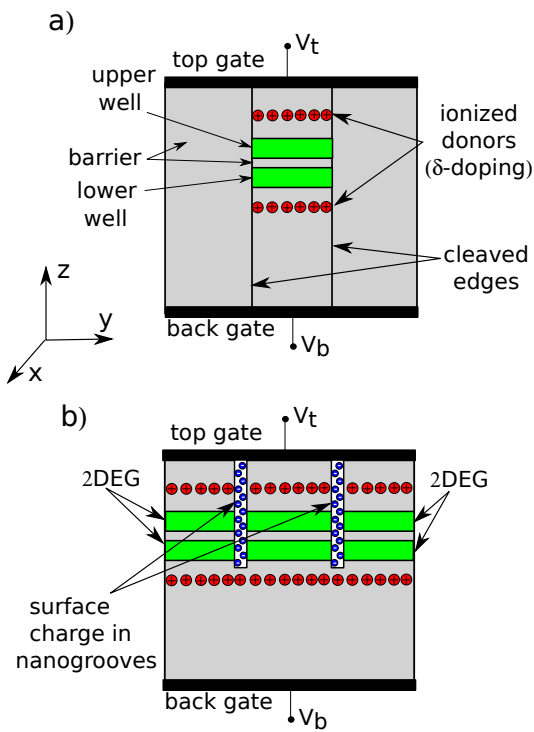}}}
        \hfill}
\caption{(Color online) Cross section of bilayer quantum wire for: a) hard wall
confinement potential as can be obtained in cleaved-edge method and b) soft lateral confinement
potential due to electrons trapped on surface of the nanogrooves created with oxidization method.}
\label{Fig:structure}
\end{figure}

The basis for fabrication of considered here two types of nanowires constitutes  a
two-dimensional layered lattice-matched
$\textrm{InP/In}_{0.52}\textrm{Al}_{0.48}\textrm{As/In}_{0.53}\textrm{Ga}_{0.47}\textrm{As}$
nanostructure prepared in molecular beam epitaxy process.
It consists of, going from bottom to top in the growth direction,  500 nm wide buffer,
 $\delta\textrm{-doping}$ (lower) donors layer, 25 nm barrier, two 15 nm wide quantum wells
separated by thin 1 nm barrier,  25 nm wide upper barrier, second $\delta\textrm{-doping}$
(upper) donors layer, 20 nm wide upper buffer and on the top, 10 nm wide capping layer.
For simplicity, we assume that all buffers and barrier layers  are made of
$\textrm{In}_{0.52}\textrm{Al}_{0.48}\textrm{As}$,
both quantum wells and top capping layer are formed within
$\textrm{In}_{0.53}\textrm{Ga}_{0.47}\textrm{As}$ regions.
The whole structure is set on n-InP substrate that is
connected to bottom back gate and another, the top gate covers the surface of the nanostructure.
Due to the heavily doped substrate, the back contact is ohmic.\cite{ohmic} However, Schottky
barrier is formed at the interface $\textrm{InP-InAlAs}$ and on $\textrm{InGaAs-top gate}$ contact.
Therefore in
calculations, we have assumed that the  lower buffer directly contacts with the back gate.
Cross-sections of both types of bilayer nanowires are shown in Fig.\ref{Fig:structure}. 
The height of Schottky barrier may differ much in the range $0.3-0.7\textrm{ V}$
for $\textrm{In}_{0.53}\textrm{Ga}_{0.47}\textrm{As}$\cite{schot_ingaas} and in the range
$0.6-0.75\textrm{ V}$ for $\textrm{In}_{0.52}\textrm{Al}_{0.48}\textrm{As}$\cite{schot_inalas}
depending on surface roughness and material of gates.
In order to remove the dependence of our results on Schottky barrier height we present them as
functions of top and bottom gates voltages  decreased by Schottky barrier i.e. $\Delta
V_{t/b}=V_{t/b}-V_{S_{t}/S_{b}}$, where $V_{S_{t}}$ and $V_{S_{b}}$ are the Schottky barriers on the
top and back contacts. In other words, $\Delta V_{t,b}$ define the shifts of conduction bands at top
and bottom interfaces.
The Fermi level in considered nanowire is defined by two external reservoirs of electrons  that is
source and drain. It is assumed to be equal $E_{F}=0$ throughout this paper. The voltages applied to
back and top gates do not change the Fermi level in the wire, which is established by source and
drain, but may change the maximal kinetic energies in both, the upper and lower layers.
We use dielectric constant $\varepsilon=14.2$.\cite{ingaas_eps}
The applied conduction band offset for InGaAs/InAlAs heterojunction equals $504\textrm{
meV}$ and the electron's effective mass $m^{*}=0.04m_{0}$.
We assumed a constant value of gyromagnetic factor as for InGaAs\cite{g_ingaas} which equals
$g^{*}=4.0$.

On InAlAs surfaces of nanogrooves the Fermi level is pinned inside an energy gap. Following Hwang
et. al.\cite{hwang} we model the density of occupied states on the barrier surface by formula
$D_{InAlAs}=D_{1}+D_{2}\cdot F_{2}$ with  $D_{1}=3\cdot 10^{11}\textrm{cm}^{-2}$ and
$D_{2}=1.61\cdot 10^{12}\textrm{cm}^{-2}$ being the densities of the lower and the upper energy
states, respectively.
The deeper surface state ($D_{1}$) is fully occupied while the occupation of the upper one ($D_{2}$)
is dependent on the position of global Fermi level and actual electrostatic potential on the
surface:
\begin{widetext}
\begin{equation}
 F_{2}=\frac{1}{\sqrt{2\pi}\sigma_{2}}\int_{-\infty}^{\infty}
 exp\left[
-\frac{\left[(E-E_{F})-(V_{b}-V_{2}+V_{Poiss})  \right]^{2}}{2\sigma_{2}^{2}}
 \right]
 f(E)dE
 \label{Eq:f2}
\end{equation}
\end{widetext}
where $E_{F}=0$ is the Fermi energy, $V_{b}=504\textrm{ meV}$ is the barrier height,
$V_{Poiss}$ is the electrostatic potential obtained as solution of Poisson equation,
$V_{2}=650\textrm{ meV}$ is the position of the upper surface state related to conduction band in
InAlAs, $\sigma_{2}=13\textrm{ meV}$ is the half-width of an upper surface state and $f(E)$ is the
Fermi-Dirac distribution function. 
Density of surface states for InGaAs is not known, but Fermi level is pinned $300\textrm{ meV}$
above the maximum of valence band.\cite{fermiingaas} We have simplified our task and assumed that
surface density is equal to $D_{InGaAs}=D_{3}\cdot F_{3}$ with $D_{3}=3\cdot
10^{11}\textrm{cm}^{-2}$ and the scaling function $F_{3}$ defined in Eq.\ref{Eq:f2} but with
quantities $V_{2}$ and $\sigma_{2}$ replaced by $V_{3}=510\textrm{ meV}$ and $\sigma_{3}=10\textrm{
meV}$.

\subsection{Calculations of electronic density and conductance}
\label{Sec:cal}

In order to calculate the density and conductance in a wire for particular set of the top and back
gates voltages we have adopted the method of lattice Green function 
described by Ando\cite{ando} and optimized by Ihnatsenka and Zozoulenko.\cite{ihnatsenka1}
First, we discretized our problem on rectangular lattice: $x_{l}=l\Delta x$, $y_{m}=m\Delta y$,
$z_{n}=n\Delta z$ with $\Delta_{\alpha}$ ($\alpha=x,y,z$) being the distances between neighbouring
mesh points in each of three dimensions. It allows us to write a tight binding Hamiltonian
for our system:
\begin{eqnarray}
 H_{x,y,z}^{\sigma}&=&\sum_{l}\sum_{m=1}^{N_{y}}\sum_{n=1}^{N_{z}}
\left\{ \right.
\left(\varepsilon_{x,y,z}+V(m,n)^{\sigma} \right)a_{lmn}^{+}a_{lmn} \nonumber \\
&-&\left( \right.
t_{x}a_{lmn}^{+}a_{l+1\ mn}e^{-I\lambda_{l,l+1}} \nonumber  \\
&+&t_{y}a_{lmn}^{+}a_{lm+1n}e^{-I\lambda_{m,m+1}}\nonumber  \\
&+&t_{z}a_{lmn}^{+}a_{lmn+1}e^{-I\lambda_{n,n+1}}
+h.c.\left. \right) \left. \right\}
\label{Eq:hxyz1}
\end{eqnarray}
In above equation we use the following notation:  $a^{+}_{lmn}$ and $a_{lmn}$ are creation and
annihilation operators defined on a lattice, $\varepsilon_{x,y,z}=2(t_{x}+t_{y}+t_{z})$ is on
site energy, $t_{\alpha}=1/2/m^{*}/(\Delta \alpha)^2$ are hopping terms for $\alpha=x,y,z$,
$V^{\sigma}(m,n)$ is an effective potential in the wire for electrons with spin $\sigma$.
Phase factors represent the Peierls shift due to vector potential $\vec{A}(\vec{r})$:
\begin{equation}
 \lambda_{p,p+1}=\frac{e}{\hbar}\int_{\vec{R}_{p}}^{\vec{R}_{p+1}}\vec{A}(\vec{r})d\vec{r}
\end{equation}
where $p=l,m,n$ defines the direction of integration when moving between two neighbouring points
($\vec{R}_{p},\vec{R}_{p+1}$) on a lattice provided that the other two indices do not change.
We choose the vector potential in nonsymmetric  gauge $\vec{A}=[zB_{y}-yB_{z},0,0]$ that gives
magnetic field $\vec{B}=[0,B_{y},B_{z}]$ which is perpendicular to direction of electron flow in a
wire.
Effective potential is given by $V^{\sigma}=V_{c}+V_{xc}^{\sigma}+g\mu B\sigma$, where $V_{c}$ is
the confinement potential in the nanostructure,
$V_{xc}^{\sigma}=\frac{\partial n^{\sigma}\varepsilon_{xc}^{\sigma}}{\partial n^{\sigma}}$ is
an exchange-correlation potential
calculated within local spin density approximation\cite{perdew}  for three dimensional
systems with Ceperley-Alder aproximation for correlation energy.\cite{ceperley} The last
term  describes the spin Zeeman factor for which we used value $\textrm{g=4.0}$ as  for
InGaAs.\cite{gfactor}

The Bloch states for free electron motion in longitudinal direction in a wire can be defined in a
mixed energy and position representation formulated by Ihnatsenka and
Zozoulenko\cite{ihnatsenka1}:
\begin{equation}
|\psi_{\alpha}^{\sigma} \rangle=\sum_{l}e^{ik_{\alpha}^{\sigma}l}\sum_{m=1}^{N_{y}}
\sum_{n=1}^{N_{z}}\psi_{\alpha}^{\sigma}(m,n)a^{+}_{lmn}|0\rangle
\label{Eq:psi}
\end{equation}
$\psi_{\alpha}^{\sigma}(m,n)$ represents a wave function for y and z quantization directions.
It can be expressed as linear combination of basis functions:
\begin{equation}
\psi_{\alpha}^{\sigma}(m,n)=\sum_{j=1}^{M} d_{\alpha,j}^{\sigma}\phi_{j}^{\sigma}(m,n) 
\end{equation}
Here we have modified slightly the method of Ihnatsenka and Zozoulenko\cite{ihnatsenka1} since
we have enabled existance of two basis for spin-up and spin-down electrons. This step is
particularily important in calculations for wire that has smooth lateral confinement
potential which is not known at the begining and a few restarts are
needed in self-consistent computations. Before each restart the recalculation of basis
functions are made for better approximated effective potentials, for spin-up and spin-down electrons
separately. Such conduct ensures size-consistency in our simulations and requires less basis
elements needed for convergence. The basis functions $\phi_{j}^{\sigma}(m,n)$ are the eigenfunctions
of two-dimensional Hamiltonian for a single slice of the wire:
\begin{eqnarray}
H_{y,z}&=&T_{yz}+V^{\sigma}\nonumber\\
&=& \sum_{m=1}^{N_{y}}\sum_{n=1}^{N_{z}}
(2t_{y}+2t_{z}+V^{\sigma}(m,n))a^{+}_{mn}a_{mn}
\nonumber \\
&-&(t_{y}a^{+}_{mn}a^{+}_{m+1n} +t_{z}a^{+}_{mn}a^{+}_{mn+1}+h.c.)
\label{Eq:hyz}
\end{eqnarray}
In case of wire with rectangular side confinement we use the same basis functions for
spin-up and spin-down electrons as obtained for pure rectangular well potential
$V^{\sigma}=V_{c}(m,n)$ whereas in case of wire with smooth side confinement we
calculate two separate basis functions for potential $V^{\sigma}=V_{c}+V_{xc}^{\sigma}+g\mu_{b}B$.
Introducing operators for creation $c^{+}_{lj}=\sum_{m,n}\phi_{j}(m,n)a^{+}_{l,m,n}$ and
annihilation $c_{l,j}=(c_{l,j}^{+})^{+}$ of basis states at any position l in a wire we may rewrite
the three-dimensional Hamiltonian in the required energy and position space representation:
\begin{eqnarray}
H_{xyz}=\sum_{l}\left\{ 
\sum_{j,j^{'}}^{M}\left(T_{jj^{'}}+V_{jj^{'}}^{\sigma}\right) c^{+}_{lj}c_{lj^{'}}
\right.\nonumber \\
\left.-\sum_{j,j^{'}}^{M}\left(
    t_{jj^{'}}^{L}c^{+}_{lj}c_{l+1j^{'}}
+ t_{jj^{'}}^{R}c^{+}_{l+1j}c_{lj^{'}}   \right)
 \right\}
 \label{Eq:hxyz}
\end{eqnarray}
The sum of kinetic and potential terms are as follows:
\begin{equation}
T_{jj^{'}}+ V^{\sigma}_{jj^{'}}=\sum_{mn}
\left[
T_{yz}(mn)+2t_{x}+V^{\sigma}(mn)
\right] \phi_{j}(mn)\phi_{j^{'}}(mn)
\end{equation}
where we skipped $\sigma$ index in basis functions while the element $T_{yz}(mn)$ is defined in
Eq.\ref{Eq:hyz}. The matrix elements $t^{R}_{jj^{'}}$ and $t^{L}_{jj^{'}}=(t^{L}_{jj^{'}})^{*}$
describe the coupling of two neighbouring slices $l$ and $l+1$:
\begin{equation}
 t^{R}_{jj^{'}}=t_{x}\sum_{mn}\phi_{j}(mn)e^{I\lambda_{ll+1}}\phi_{j^{'}}(mn)
\end{equation}
Further procedure is identical as in the method of Ihnatsenka and Zozoulenko for strictly
two-dimensional quantum wire\cite{ihnatsenka1}, that is, we calculate Green
function for a single slice, determine the Bloch states $\psi_{\alpha}^{\sigma}$ and wave vectors
$k_{\alpha}^{\sigma}$ for propagating and evanescent modes for given energy. Having wave
functions $\psi_{\alpha}^{\sigma}$ we used them to construct the surface Green functions of the
left ($\Gamma^{L,\sigma}$) and right ($\Gamma^{R,\sigma}$) semi-infinite wires which allows finally
to get the Green function $G^{\sigma}(mn,m'n',E)$ for the infinite bi-layer wire. For more details
see work [\cite{ihnatsenka1}].
Having the Green function for infinite wire,  we use a standard formula for local density of
states $\rho^{\sigma}(mn,E)=-Im[G^{\sigma}(mn,mn,E)]/\pi$ and the total spin electronic density
$n^{\sigma}_{e}(nm)=\int_{E_{min}}^{\infty}\rho^{\sigma}(nm,E)f(E) dE$. 
The total charge density in a system is the sum of electron density confined in the lower and upper
quantum wells  ($n_{e}^{\sigma}$), charge of ionized donors in the lower ($n_{1}$) and in the upper
($n_{2}$) $\delta\textrm{-doped}$ layers and for second type wire only, the surface charge gathered
in nanogroves ($n_{s}$):
\begin{equation}
n_{tot}=n^{\uparrow}_{e}+n^{\downarrow}_{e}+n_{s}-n_{1}-n_{2}
\end{equation}
We use this charge density in Poisson equation which is solved with von Neumann boundary
conditions on left and right sides of the nanostructure while Dirichlet boundary conditions were
used on top and bottom gates. For particular set of top and back gate voltages, 
the total charge density and the effective potential for considered type of lateral confinement
were computed iteratively in self-consistent manner. Iterations were continued until relative change
in total denisty was less than $10^{-4}$ what also have guaranteed convergance of effective
potential.
Then, the value of conductance was determined according to Landauer formula:
\begin{equation}
G^{\sigma}_{cond}=\frac{e^2}{h}\sum_{\alpha}\int_{E_{F}-6kT}^{E_{F}+6kT}
\left(-\frac{df}{dE}\right) dE
\end{equation}
where  $\alpha$ indicates summation over all active transport channels for given Fermi energy.
Here, we are particularly intereseted in spin polarization of a conductance.
For quantitative analysis we used the following definition of spin polarization of considered
quantity: $\eta_{X}=(X^{\uparrow}-X^{\downarrow})/(X^{\uparrow}+X^{\downarrow})$, where
$X^{\uparrow}$ and $X^{\downarrow}$ correspond to spin-up and spin-down parts of that quantity
which in our work are: the conductance $G_{cond}$ and the total electronic density $n_{e}$ in the
nanowire.

The specific properties of a bi-layer wire that appear in the transverse high magnetic field depends
strongly on the energy splitting between the lowest symmetric and antisymmetric states $\Delta
E_{SAS}$.\cite{fischer}
For low values  of $\Delta E_{SAS}$, the effects of vertical modes mixing get stronger, that is,
there are developed deeper side minima in energy dispersion relation E(k).\cite{chwiej-bilayer}
Therefore, in order to monitor this quantity, for each work point i.e. the  pair of confinement
potential shifts $\Delta V_{b}$ and $\Delta V_{t}$,  after a self-consistency was reached, we have
also determined the lowest eigenenergy  and the corresponding wave function for isolated, the lower
and upper wells.
Diagonalization of  Hamiltonian given by Eq.\ref{Eq:hyz} with this two member basis
$\{\psi_{l},\psi_{u}\}$ provides required energies for symmetric ($E_{S}$) and antysymmetric
($E_{AS}$) two-wells state provided that both wells have more or less the same shape and depth. In
such case $\Delta E_{SAS}=E_{AS}-E_{S}$ gets its minimum. Otherwise, due to detuning of both
single-well states, the value of $\Delta E_{SAS}$ grows.

\section{Results}
\subsection{Bi-layer nanowire with rectangular lateral confinement potential}
\label{Sec:rect}
\begin{figure}[htbp!]
\hbox{
	\epsfxsize=80mm
       \rotatebox{0}{{ \epsfbox[10 310 390 830] {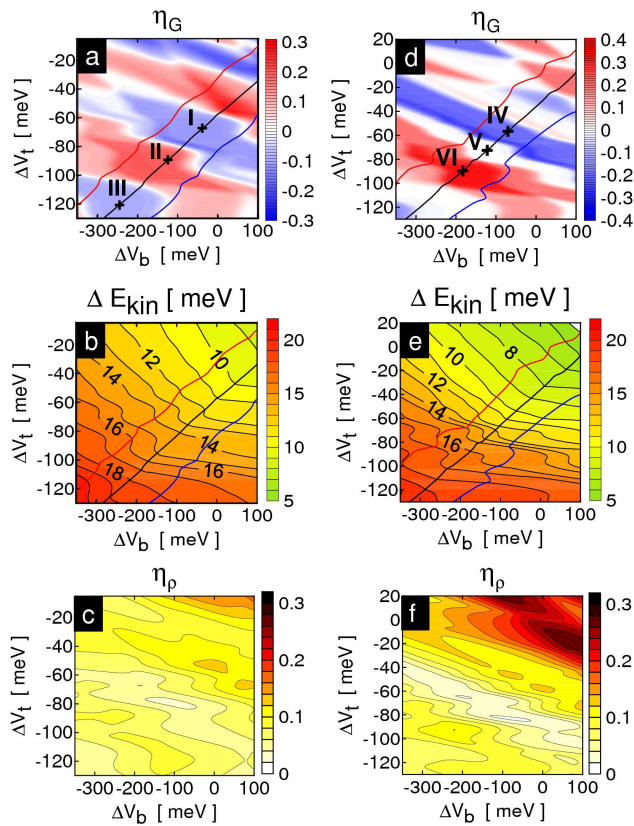}}}
        \hfill}
\caption{(Color online) (a, d) Spin polarization of conductance, (b, e) maximum of kinetic
energy and (c, f) spin polarization of total electronic density for bi-layer nanowire with
rectangular lateral confinement potential and for two widths of wire:
$y_{w}=120\textrm{ nm}$ (left column) and $y_{w}=60\ \textrm{nm}$ (right column).
Red, black and blue lines
in first row indicate isolines for energy difference between the lowest energy
states in upper (u) and in lower (l) wells $\Delta E_{ul}=1,0,-1\ \textrm{meV}$,
respectively. Crosses numbered from I to VI  are work points ($\Delta
V_{b},\Delta V_{t}$) which are analyzed in text. For work points, the energy splitting
$\Delta E_{SAS}$ gets the following values: (I) $4.9\textrm{ meV}$, (II) $4.8\textrm{ meV}$, (III)
$4.4\textrm{ meV}$ and (IV) $5.1\textrm{ meV}$, (V) $5.0\textrm{ meV}$, (VI) $4.9\textrm{ meV}$.}
\label{Fig:edpol}
\end{figure}

In calculations performed for rectangular confinement potential we used following
parameters:
$\textrm{B}_{y}=10\textrm{ T}$, $\textrm{B}_{z}=1\textrm{ T}$,
$\Delta_{x}=\Delta_{y}=1\textrm{ nm}$, $\Delta_{z}=0.5\textrm{ nm}$,
densities of dopants in lower and in upper $\delta$-layers were equal $\rho_{1}=3\cdot
10^{11}\textrm{cm}^{-2}$ and $\rho_{2}=4\cdot 10^{11}\textrm{cm}^{-2}$ while
the ionization energy of dopants was set to $5\textrm{ meV}$. All results in this and in the next
section were obtained for  temperature $\textrm{T=1 K}$. 

\begin{figure}[htbp!]
\hbox{
	\epsfxsize=80mm
       \rotatebox{0}{{ \epsfbox[0 318 383 842] {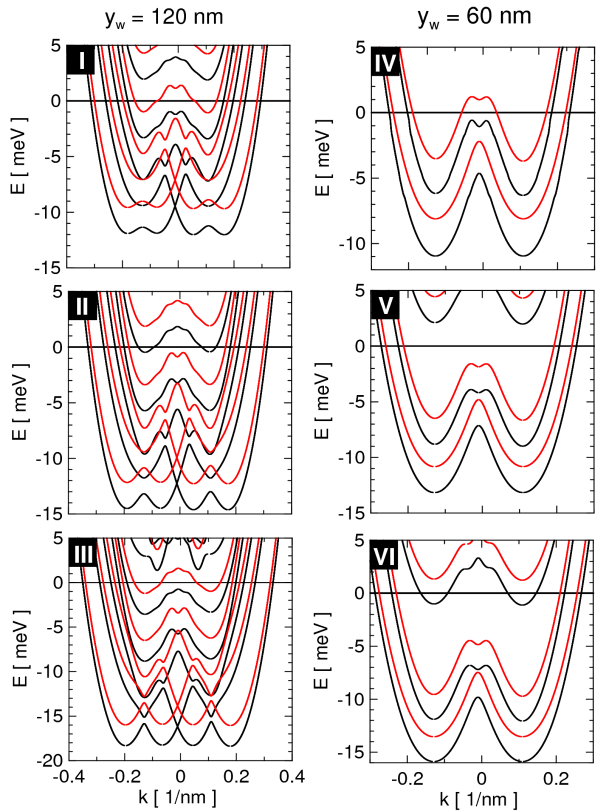}}}
        \hfill}
\caption{(Color online) Energies of spin-up (black color) and spin-down (red color)
magnetosubbands for bi-layer quantum wire of width $y_{w}=120\textrm{ nm}$ (left column) and
$y_{w}=60\textrm{ nm}$ (right column) with rectangular confinement potential.
Horizontal black line marks Fermi level in the system. Numbers  $I-VI$ correspond to
work points marked in Figs. \ref{Fig:edpol}(a) and \ref{Fig:edpol}(d).}
\label{Fig:edek1}
\end{figure}

Figure \ref{Fig:edpol} show spin polarization of conductance ($\eta_{G}$), maximal kinetic
energy ($\Delta E_{kin}$)  and spin polarization of total electronic density ($\eta_{\rho}$) for
bi-layer nanowire of $y_{w}=120\textrm{ nm}$ and $y_{w}=60\textrm{ nm}$ widths.
For wider wire, polarization of conductance depends on both $\Delta V_{b}$
and $\Delta V_{t}$ values but maximal value of $\eta_{G}$ does not exceed $30\%$
[Fig.\ref{Fig:edpol}(a)].
Simultaneous change in top and back gates voltages along a black line,
which marks the work points for which lowest energy levels in upper and in lower wells are equal
($\Delta E_{ul}\approx 0$), makes $\eta_{G}$ to oscillate. Sign of $\eta_{G}$ is changed when
next subband come into a transport window and due to spin Zeeman energy splitting of spin-up and
spin-down $E(k)$ branches [Fig.\ref{Fig:edek1}], positive and negative stripes lie alternately
being separated by very thin unpolarized regions.
What is interesting, when $\Delta E_{ul}$ is detuned and gets value of $\pm 1\textrm{ meV}$ [red and
blue line in Fig.\ref{Fig:edpol}], pattern of $\eta_{G}$ does not change much besides
small distortions.
For the range of analyzed here  values of $\Delta V_{b}$ and $\Delta V_{t}$, maximal
kinetic energy may change bewteen 9 meV and 20 meV [Fig.\ref{Fig:edpol}(b)]
with number of active subbands between 3 and 5.
In our earlier work\cite{chwiej-bilayer} we have found that for similar range of kinetic energy,
spin polarization of conductance may achieve even $60\%$. However, those preliminary calculations
were performed neglecting electrostatic interactions between electrons.
In considered range of kinetic energy, polaritzation of total electronic density is low and
for $\Delta E_{kin}>10\textrm{ meV}$ its value remains lower than $10\%$ close to the line $\Delta
E_{ul}\approx
0$. Small polarization of electron density is due to small energy splitting of
spin-up and spin-down energy subbands which come into transport window alternately[see
Fig.\ref{Fig:edek1}].

\begin{figure}[htbp!]
\hbox{
	\epsfxsize=80mm
       \rotatebox{0}{{ \epsfbox[0 318 595 842] {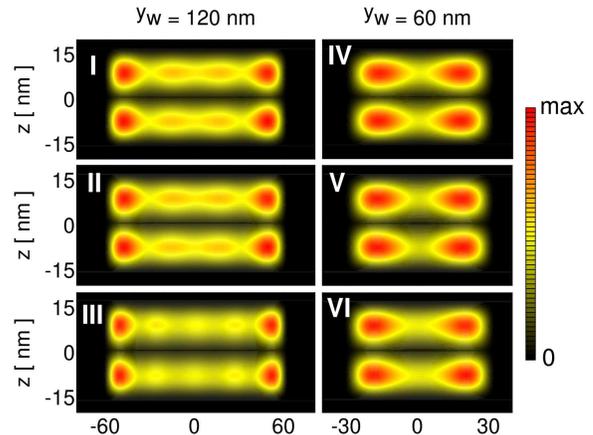}}}
        \hfill}
\caption{(Color online) Electron densities in bi-layer quantum wire of width $y_{w}=120\textrm{ nm}$
(left column) and $y_{w}=60\textrm{ nm}$ (right column) with rectangular lateral confinement
potential.
Roman numbers ($\textrm{I-VI}$) correspond to work points marked in Figs. \ref{Fig:edpol}(a) and
\ref{Fig:edpol}(d).}
\label{Fig:edro}
\end{figure}
Such large divergance of results obtained within non-interacting model\cite{chwiej-bilayer} and in
the present one has forced us to analyzed this problem in
detail. In figure \ref{Fig:edpol}(a) we have marked three arbitrarily chosen work points and number
them I, II and III.
Total electron densities calculated for these work points are presented in Fig.\ref{Fig:edro}.
We notice that in each well there are developed two maximums, in the  left and in the right corners.
Electrons confined in wire strongly repeals each other. When they are trying to avoid other
electrons they mostly localize at the edges of the system due to a lower electrostatic potential.
For points I and II in Fig.\ref{Fig:edro} there are two additional smaller maixmums located near the
center of nanowire. When total electron density becomes larger (point III), three
hardly seen smaller local maximums instead of two are developed. Similar formation of edge
channels due to electron-electron interaction was predicted also for a single-layer
nanowire.\cite{ihnatsenka2}
In order to check, if the formation of edge wells can be a reason of low polarization of
conductance we have repeated calculations for thinner wire which was $y_{w}=60\textrm{ nm}$ wide.
Indeed, for two times thinner wire, the amplitude of conductance polarization grows up to
$40\%$ [Fig.\ref{Fig:edpol}(d)]. Moreover, the patterns of $\eta_{g}$ for wide and narrow wire
differ qualitatively. The stripes for partly spin-up (red) and spin-down (blue) polarization are
narrower with sharp edges and the unpolarized regions get wider for narrower wire.  The isopotential
lines in $\Delta E_{kin}$
picture are densier in the middle part of figure \ref{Fig:edpol}(e) and are well
correlated with the region of unpolarized total density depicted in Fig.\ref{Fig:edpol}(f). Due
to small amount of active transport channels ($\le 2$) for $\Delta E_{kin}$ values less than
$8\textrm{ meV}$  polarization of total density can be as large as $30\%$.
However, the decrease of wire width has not allowed for elimination of lateral
edge wells what confirms the electron densities displayed in second
column of Fig.\ref{Fig:edro} for work points: IV, V and VI which correspond to those  marked
in  Fig.\ref{Fig:edpol}(d).
Although for both, wide and narrow wires, maximums of density are located at the edges, their
energy subbands differ much. Energies of magnetosubbands for points I-III
($y_{w}=120\textrm{ nm}$) and points IV-VI ($y_{w}=60\textrm{ nm}$) are shown in
Fig.\ref{Fig:edek1}.
When  wire is narrow (second column), subbands have two distinct lateral minima. Two
lowest subbands are separated from the upper ones what explains an appearance of large
unpolarized region in $\eta_{G}$ and in $\eta_{\rho}$ [cf. Figs. \ref{Fig:edpol}(d) and
\ref{Fig:edpol}(f)]. The shapes of magnetosubbands for narrower bi-layer wire is preserved
when potentials $\Delta V_{b}$ and $\Delta V_{t}$ are decreased along the line for $\Delta
E_{ul}\approx
0$. Partial polarization of conductance appears when number of Fermi level crossings
with spin-up (spin-down) energy branches is larger than those for spin-down (spin-up) 
ones. It is easily noticable in Fig.\ref{Fig:edek1} for points IV and VI while for work point V
the conductance is unpolarized.

Magnetosubbands for wider wire have more complicated shapes and the energy spacings between them are
smaller than for narrow wire. Moreover, when depth of effective confinement potential is
lowered along the line $\Delta E_{ul}\approx 0$ the shapes of few lowest subbands undergo
significant modifications. The depth of two weakly marked lateral minimums in lowest subband on both
sides of $k=0$ for point I grows twice when work point is shifted to point III [cf.
Figs. \ref{Fig:edek1}(I) and \ref{Fig:edek1}(III)].
During this process the lateral minimums of second subband, for $k\ne 0$, approach the maximums of
the lowest subband what results in their stronger mixing.
For this reason, smooth so far the energy minima and maxima now are replaced by cusps [see
lower part of spin-up, first and second subbands in Fig.\ref{Fig:edek1}(III)].
In addition, the large amplitudes of energy oscillations in vicinity of $k=0$ make spin-up
and spin-down branches to be not well separated what is the main condition for appearence of 
large spin polarization of conductance in bi-layer system.\cite{chwiej-bilayer}
We may thus conclude that a large spin polarization of conductance in bi-layer
nanowire with hard lateral confinement can not be reached due to significant modifications of
confinement potential as well as shapes of magnetosubbands.

An appearance of double lateral minumums in lowest subband can be explained in semiclassical model.
For vector potential used in this paper $\vec{A}=[zB_{y}-yB_{z},0,0]$, the x component of 
canonical wave vector ($\vec{k}=(\vec{p}+e\vec{A})/\hbar$) has the following form:
$k_{x}=e(zB_{y}-yB_{z})/\hbar$. The difference of $k_{x}$ components of wave vectors of two
electrons which move in the left and in the right edge wells lying within the same layer is
$\Delta k_{x}=e\Delta y B_{z}/\hbar$. For average distance between centers of both subwells [see
Fig.\ref{Fig:edro}] being equal $\Delta y=100\textrm{ nm}$ it gives wave vector shift
$\Delta k_{x}=0.152\textrm{ nm}^{-1}$. This value agrees quite well with the difference
between two lateral minimums in lowest subband in Fig.\ref{Fig:edek1}(III) for $k>0$ or $k<0$.
\begin{figure}[htbp!]
\hbox{
	\epsfxsize=80mm
       \rotatebox{0}{{ \epsfbox[0 313 342 842] {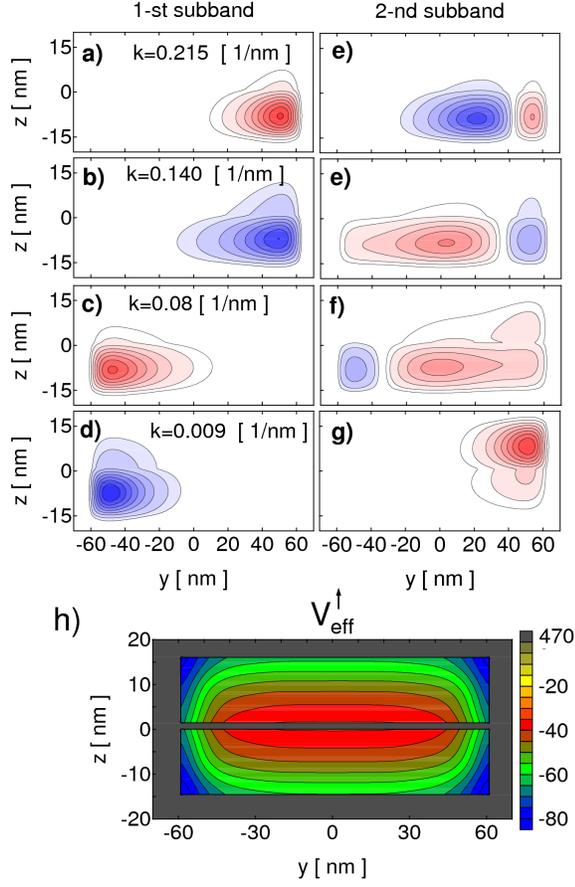}}}
        \hfill}
\caption{(Color online) Wave functions $\psi_{k}^{\uparrow}(y,z)$ for first (a-d) and second
(e-g) subbands calculated for rectangular confinement potential of width $y_{w}=120\textrm{
nm}$ and kinetic energy $\textrm{E}_{kin}=1\textrm{ meV}$ ($k>0$).
Wave functions are related to magnetosubbands shown in Fig.\ref{Fig:edek1}(III).
The lowest row (h) shows a cross-section of relaxed effective potential.
Wave functions and effective potential were calculated for spin-up electrons.
}
\label{Fig:ekpsi}
\end{figure}
For narrower wire, $\Delta k_{x}$  has two times smaller value but due to the larger energy spacings
between subsequent lateral modes [compare large energy spacings between first and second 
subbands in Fig.\ref{Fig:edek1}(VI) with their smaller counterparts shown in
Fig.\ref{Fig:edek1}(III)] mixing of lateral modes is less effective what prevents the formation of
two additional minimums in the lowest subband.

Splitting of the current flow into two parralel paths becomes also apparent if we look 
at the wave functions of first subband for $k>0$ and kinetic energy $E_{kin}=1\textrm{ meV}$ (total
energy is $\textrm{E= -17.4 meV}$) which are displayed in Fig.\ref{Fig:ekpsi}.
To get the wave function within our model, we first calculated the wave vectors
$k^{\sigma}_{\alpha}$ for given energy and next used them to reduce the dimensionality of
Hamiltonian given in Eq.\ref{Eq:hxyz1}:
\begin{equation}
H_{yz}(k_{\alpha}^{\sigma})= \langle e^{ik^{\sigma}_{\alpha}}
|H_{x,y,z}^{\sigma}|\psi_{\alpha}^{\sigma}\rangle
\end{equation}
with wave functions $|\psi_{\alpha}^{\sigma}\rangle$ defined in Eq.\ref{Eq:psi}. Finally, the  
effective two-dimensional Hamiltonian for known $k_{\alpha}^{\sigma}$ has the following form:
\begin{equation}
 H_{yz}(k_{\alpha}^{\sigma})=H_{yz}
 +\sum_{mn}\frac{1-cos[k_{\alpha}^{\sigma}\Delta x(z\omega_{y}-y\omega_{z})]}{m^{*}(\Delta x)^2}
 a_{mn}^{+}a_{mn}
\end{equation}
where $H_{yz}$ is an energy operator for single slice [Eq.\ref{Eq:hyz}] while $\omega_{y}$ and
$\omega_{z}$ are the cyclotron frequencies for y and z components of $\vec{B}$. The wave functions
for first and second subbands which were obtained during diagonalization of
$H_{yz}(k_{\alpha}^{\sigma})$ are shown in figures \ref{Fig:ekpsi}(a-g).
All wave functions for first subband are localized in lower layer even though
half of these states have negative values of group velocity $v=(\partial E/\partial k)/m^{*}$
[states (b) and (d)]. At first, one may think that the lowest-subband states with negative velocity
should always lie in the upper layer due to action of strong magnetic force in vertical direction
induced by large value of $B_{y}=10\textrm{ T}$. However, the effective potential in both layers is
not the same.
Since densities of dopants in lower and in upper $\delta$-layers differ much and due to the
fact that the distance between the top gate and the upper well as well as the distance from the
lower well to the back gate are different, tuning of gates voltages can not equalize the
effective potentials in both layers. That is also the reason of small negative k shift in all
dispersion branches shown in Figs.\ref{Fig:edek1} even though the depth of energy minima for $k>0$
and $k<0$ are at the same level.
Thus,  localization of wave functions for $k>0$ or $k<0$ for lowest subband is in slightly deeper
well, independently on the direction of actual (small) group velocity.
In agreement with our semiclassical explanation of quadruple energy minimums formation,
one pair of wavefunctions belonging to the right most minimum [Figs.\ref{Fig:ekpsi}(a,b)] are
entirely localized in the lower right well while a pair of wavefunctions for second energy minimum
which has lower k occupy the left one [Figs.\ref{Fig:ekpsi}(c,d)].
In Fig.\ref{Fig:ekpsi}(e-f) we see that the wave functions even for second subband are pinned to
particular, left or right well, although these states extend over a larger area
than those for first subband. For narrower wire, wave function for second subband extend more
evenly over a single layer, lower or upper, what eventually prevents formation of separate
energy minimums for the left and for the right edge wells.

\subsection{Nanowire with smooth lateral confinement potential}
\label{Sec:smooth}

In previous subsection we have shown that the spin polarization of conductance for bi-layer nanowire
with rectangular lateral confinement potential can not reach more than $40\%$. It results from the
fact that formation of two lateral  edge wells make an electronic density to be nonhomogeneous
what effectively distorts energy subbands. Therefore we repeated calculations for smooth
confinement potential as can be fabricated by means of oxidization method and which is supposed to
form more homogeneous electron density and in consequence shall give less distorted magnetosubbands.
Calculations were performed for two nanogrooves depths:  85 nm and 115 nm.
In first case, the widths of upper and lower quantum wells formed in both layers are different [see
extent of
electron density in nanowire in Figs.\ref{Fig:edro2}(I-III)] since bottom of lower
well lies below that of nanogroove and thus  the surface charge gathered in nanogooves has less
impact on lower layer than on an upper one. In the latter case, the  widths of the upper and lower
quantum wells become comparable [Figs.\ref{Fig:edro2}(IV-VI)] what may influence the results.
We assume that single nanogroove is $y_{g}=40\text{ nm}$ wide\cite{groove_width} and the
surface states are located at its both sides.
Distance between centers of nanogrooves equals $270\textrm{ nm}$. On both sides of
the nanowire in lateral direction a two-dimensional electron gas is confined in 500 nm wide
channels [see Fig.\ref{Fig:structure}(b)]. Density of 2DEG was calculated similarily as for nanowire
with an exception that von Neumann boundary conditions were imposed on the left and on the right
outermost boundaries of wave
functions constituting an electron gas.
To determine the surface charge gathered on nanogroove sides we have used phenomenological
model provided by Hwang et. al. in work[\cite{hwang}] and briefly described in Sec.\ref{Sec:model}.
We also diversified densities of dopants in $\delta-$layers
to make the electron densities confined in upper and in lower quantum wells comparable.
For nanogrooves depth $h=85\textrm{ nm}$ we set densities of donors to $\rho_{1}=1.2\cdot
10^{11}\textrm{cm}^{-2}$ (lower well) and $\rho_{2}=2.8\cdot 10^{11}\textrm{cm}^{-2}$ (upper well)
while for deeper nanogrooves ($\textrm{h=115 nm}$) we increase density of dopants in lower layer to
$\rho_{1}=1.8\cdot 10^{11}\textrm{cm}^{-2}$ and leave $\rho_{2}$ unchanged. 
\begin{figure}[htbp!]
\hbox{
	\epsfxsize=80mm
       \rotatebox{0}{{ \epsfbox[10 290 390 830] {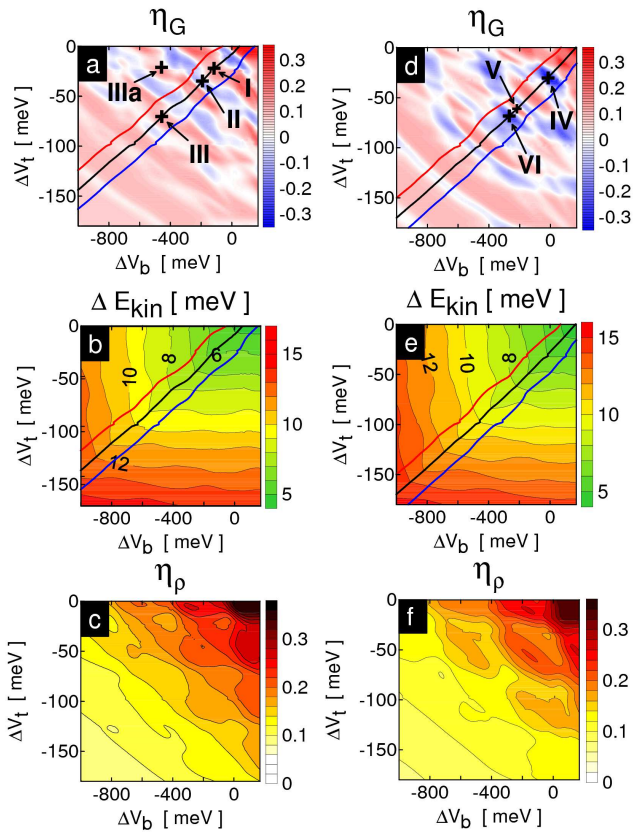}}}
        \hfill}
\caption{(Color online) (a,d) Spin polarization of conductance, (b,e) maximum of kinetic
energy and (c,f) spin polarization of total electron density for bi-layer nanowire with smooth
lateral confinement potential. Depths of nanogrooves are: $\textrm{85 nm}$ (left column) and
$\textrm{115 nm}$ (right column). The red, black and blue lines in
first row indicate isolines for energy difference between the lowest energy
states in the lower and in the upper wells $\Delta E_{ul}=1,0,-1\ \textrm{meV}$, respectively.
Numbers $\textrm{(I-VI)}$ marks the work points ($\Delta V_{b},\Delta V_{t}$) which are analyzed in
text.}
\label{Fig:edpol2}
\end{figure}
\begin{figure}[htbp!]
\hbox{
	\epsfxsize=80mm
       \rotatebox{0}{{ \epsfbox[0 370 595 842] {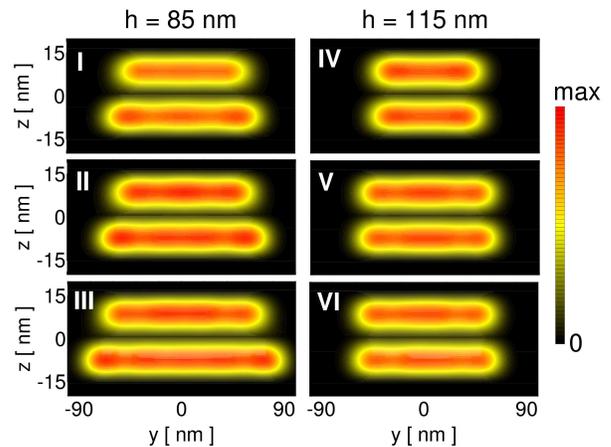}}}
        \hfill}
\caption{(Color online) Electron densities in bi-layer quantum wire with smooth lateral confinement
potential. Nanogrooves widths are: 85 nm (left column) and 115 nm (right column).
Roman numbers ($\textrm{I-VI}$) correspond to work points marked in
Figs.\ref{Fig:edpol2}(a,d).}
\label{Fig:edro2}
\end{figure}
\begin{figure}[htbp!]
\hbox{
	\epsfxsize=80mm
       \rotatebox{0}{{ \epsfbox[0 133 379 842] {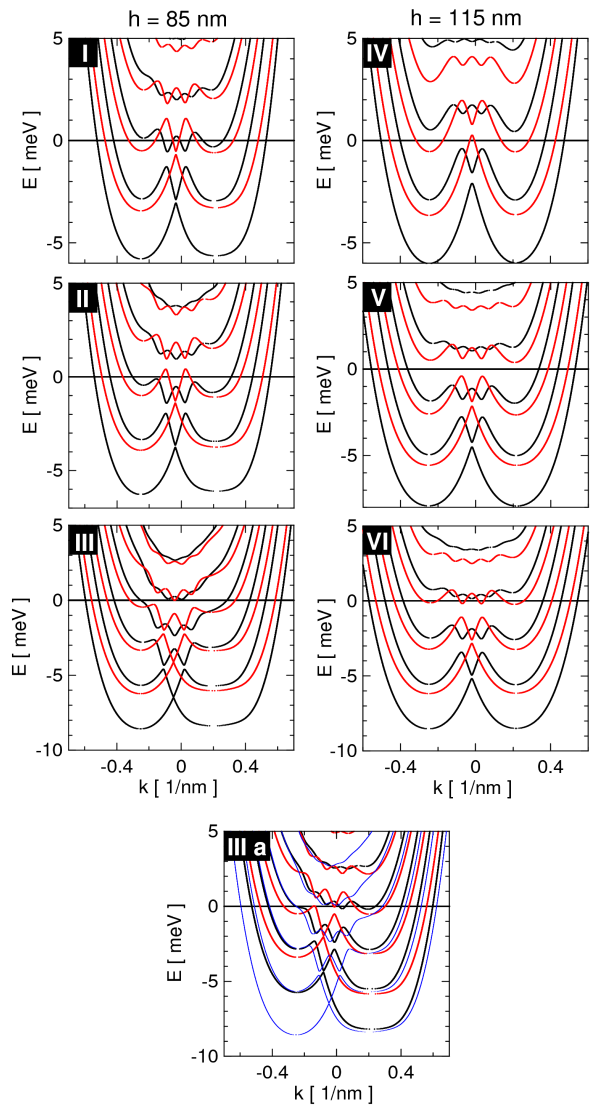}}}
        \hfill}
\caption{(Color online)
(Color online) Spin-up (black color) and spin-down (red color) energy dispersion
for the bilayer quantum wire soft wall confinement potential.
Horizontal black line marks the Fermi level in the system.
Numbers I, II, III and IIIa correspond to work points marked in Fig.\ref{Fig:edpol2}(a) while
numbers IV, V and VI correspond to those on Fig.\ref{Fig:edpol2}(d).
Blue color energy branches in case IIIa are the copies of spin-up (black) branches taken from work
point III for comparison only.}
\label{Fig:edek2}
\end{figure}

The spin polarization of conductance and electron density confined in bi-layer nanowire as
well as maximal kinetic energy as function of $\Delta V_{b}$ and $\Delta V_{t}$ are shown in
Fig.\ref{Fig:edpol2}.
We notice that values of $\eta_{G}$  do not exceed $33\%$ for both nanogroove depths.
It is not however an effect of granularity of electron density.
Densities shown in Fig.\ref{Fig:edro2} for arbitrarily chosen three work points have in both cases 
distinct maximums which are smooth and elongated in y direction.
An unexpected low spin polarization of conductance can not be also an effect of detuning
the energy ladders in both quantum wells since densitites that have
been obtained for deeper nanogroves are almost the same in both layers what excludes such
possibility.

It is interesting that the lines of constant kinetic energy $\Delta E_{kin}$ in
Fig.\ref{Fig:edek2}(b,e) resemble a rectangle corner.
When value of $\Delta V_{b}$ ($\Delta V_{t}$)  is fixed and $\Delta V_{t}$ ($\Delta V_{b}$) is
increased then the value of kinetic energy varies slightly.
It results from fact that upper and lower quantum wells are strongly coupled only to one, 
top or back gate. For example, the change of top gate voltage  from work point
III to IIIa in Fig.\ref{Fig:edpol2}(a) make an energy shift in these parts of magnetosubbands which
describe an electron motion in upper layer [left part of energy spectrum in
Fig.\ref{Fig:edek2}(IIIa)] leaving simultaneously an electron motion in lower layer unchanged [right
part in Fig.\ref{Fig:edek2}(IIIa)]. This strong and selective coupling results from screening
properties of an electron gas which is confined at both sides of nanowire
[see Fig.\ref{Fig:structure}(b)]. Variations of voltages applied to top (back) gate make an
upper quantum well deeper or shallower what implies additional charge flow between an upper
(lower) layer and external reservoirs of electrons.
This effect was insignificant in nanowire with rectangular confinement
potential since the electron density gathered only in nanowire can not screen much wider metallic
gates. Therefore, variations of kinetic energy in Figs. \ref{Fig:edpol}(b) and \ref{Fig:edpol}(e)
have such irregular patterns unlike these visible in Figs. \ref{Fig:edpol2}(b) and
\ref{Fig:edpol2}(e).

Energy spectra calculated for work points I-VI lying on a line $\Delta E_{ul}\approx 0$ and marked
in Fig.\ref{Fig:edpol2}(a,d) are presented in Fig.\ref{Fig:edek2}. For nanogrooves depth
$\textrm{h=85 nm}$, the lower layer in nanowire is not completely surrounded by nanogrooves
as it is for the upper layer. For this reason, the energy branches for $k<0$  differ from these
for $k>0$. In Fig.\ref{Fig:edek2}(I-III) we notice, that energy branches describing
mainly the motion of electrons in the lower well (right part) are wider than these for the upper one
(left part). For deeper nanogooves ($h=115\ nm$) both nanowire layers interacts electrostatically
with the surface charge in similar manner. The widths of the upper and lower wells are the same
[Fig.\ref{Fig:edro2}(IV-VI)] what translates into almost perfect mirror symmetry of energy subbands.
Independently of the depth of nanogrooves, the three lowest spin-down branches (red color) are
shifted up on energy scale due to spin Zeeman effect but localize near subsequent upper spin-up
subbands (black color). This is a main reason of low spin polarization of conductance in nanowire
with smooth confinement potential.
Although, the amplitudes of oscillations in $E(k)$ in the vicinity of  $k=0$ seems to be optimal for
high polarization of conductance\cite{chwiej-bilayer} because the energy oscillations in two
neighbouring subbands are separated. But actually due to frequent overlapping of opposite spin
magnetosubbands at Fermi energy level, spin polarization stays low.
For the same reason, polarization of the total electronic density confined in bi-layer wire is 
low besides a region with small values of $\Delta V_{b}$  and $\Delta V_{t}$  [upper right part in
Fig.\ref{Fig:edpol2}(c,f)] when only two spin-up and one spin-down subbands cross Fermi 
level.

Value of gyromagnetic factor in InGaAs material that constitutes quantum wells is dependent on
voltages applied to gates\cite{gfactor} what is omitted in our model. But, if we reasonably assume,
it may change in the range $g=3.5-4.2$ then the energy shift due to spin Zeeman
splitting\cite{splitting} varies between $2.03\textrm{ meV}$ and $2.43\textrm{ meV}$ (for $g=4$ is
equal $2.31\textrm{ meV}$). Thus, the variations of g values induced by the changes of
voltages applied to the metallic gates are too small to significantly draw aside the spin-up and
spin-down subbands shown in Fig.\ref{Fig:edek2}(I-VI). It seems, the only way to get a large  spin
polarization of conductance is to tune the magnetic field, especially value of $B_{y}$ as it gives
the main contribution to B. However, the change of magnetic field may distort much the shapes of
magnetosubbands and in such case an additional tuning of both, top and back gate voltages could
be required.

\section{Conclusions}
\label{Sec:con}

We have calculated the conductance of bi-layer nanowire in tilted magnetic field taking into
account an electrostatic interaction between electrons in nanosystem and for two types of lateral
confinement potential: smooth and rectangular. Our considerations have mainly been focused on spin
polarization of conductance which can achieve even $60\%$ for moderate Fermi energy if interaction
is neglected.
Simulations were performed for wide range of voltages applied to top and back gates which allow to
change  the Fermi energy in lower and in upper quantum well. 
It was shown that in nanowire with rectangular lateral confinement potential,
maxima of electron density are located at side edges of wire.
If nanowire is wide then the lowest subband and the main part of second subband are localized in the
same edge well. In such case, there appear two additional energy minimums in lowest magnetosubband.
Independenlty of width of rectangular  bi-layer wire, pseudogaps are opened in energy
spectrum but due to large distortions of magnetosubbands only a moderate ($\le 40\%$) spin
polarization of conductance can be reached.
If lateral confinement potential in wire is smooth, local maxima in electron density are
diminished and since the edge states dissappear, magnetosubbands are less distorted.
Even though, the widths of pseudogaps are optimal for large spin polarization, in some cases it can
not be reached due to an overlap of spin-up and spin-down magnetosubbands. Such overlap is
accidental and results from spin Zeeman energy shift of spin-up and spin-down energy subbands.
We may thus conclude, that in bi-layer nanowire with smooth lateral confinement potential, the
large spin polarization of conductance can be achieved only if both, the magnetic field and the
Fermi energies in upper and lower layers of wire will be simultaneously tuned.

\section*{Acknowledgements}
\noindent
The work was financed by Polish Ministry of Science and Higher Education (MNiSW)


\begin{thebibliography}{00}
\bibitem{splitting} Another contribution to energy shift of spin-up an spin-down subbands gives the
exchange-correlation potential. This type of energy splitting also depends on gyromagnetic value
since the larger g is, the larger difference between spin-up and spin-down densities we get.
\end{thebibliography}
\end{document}